\documentclass[11pt,a4paper]{article}
\usepackage{epsf}

\def\hc#1{\leavevmode\hbox to \hsize{\hss #1\hss}\leavevmode}

\begin{document}


\title{Measuring the UHE cosmic-ray composition\\
 with tracking detectors in air shower arrays}
\author{Konrad Bernl\"ohr \\
 {\normalsize\it Max-Planck-Institut f\"ur Kernphysik,} \\
  {\normalsize\it Postfach 103980, D-69029 Heidelberg, Germany}}
\date{January 16, 1996}

\maketitle

\begin{abstract}

Measuring the angles of muons and electrons in air showers is proposed as a
method for studying the primary cosmic-ray mass composition near 
the knee of the cosmic-ray energy spectrum at a few $10^{15}$\,eV.
Conventional tracking detectors at existing air shower arrays could serve
this purpose, like the CRT detectors at the HEGRA array.
When the average radial muon angles are examined as a function of shower
core distance, the experimental resolution can be very well calibrated
from the tangential angle distribution. The method is particularly
promising for measuring changes in the average mass number of the
primary cosmic rays with energy. The method is described and
experimental and theoretical constraints are discussed.

\end{abstract}


\section{Introduction}

Despite the fact that ultra-high energy (UHE) cosmic rays are known
for decades, their sources and  the acceleration mechanisms are
still under debate. Sources are only detectable by $\gamma$-rays produced
in interactions near the sources.
In the very-high energy (VHE) range near 1\,TeV 
more and more $\gamma$-ray sources are revealed by the imaging Cherenkov technique.
On the contrary, no clear source detections have been made in the UHE domain
above about 100\,TeV, except perhaps a few episodic cases. 

Mainly for reasons of the required power, the dominant sources of
cosmic rays up to about 100\,TeV and probably up to the {\em knee}
of the cosmic-ray energy spectrum at a few $10^{15}$\,eV are believed
to be supernova remnants in the Sedov phase. The change of the spectrum
near the knee presumably reflects a change in the origin and the
takeover of another, yet unclear type of sources at energies above
the knee. A change in the cosmic-ray propagation with a decreasing
Galactic containment has also been considered. In either case,
the change in the slope of the spectrum should be accompanied by a change 
in the mass composition of cosmic rays. In the case of a change of
sources across the knee, the composition could change dramatically.
A much less spectacular change of the composition is expected
in the case of decreasing containment.

Direct measurements well below the knee
\cite{Asakimori-1993ab,Ichimura-1993b}
show indeed a substantial change in the mass composition
already at energies around 100\,TeV. In particular, the fraction
of protons seems to diminish with increasing energy.
Due to their small collection
areas the balloon-borne direct experiments run out of statistics
above several hundred TeV. Indirect methods using ground-based experiments
are, so far, not able to classify individual cosmic rays unambiguously
by their mass. Such methods are more appropriate for evaluating some
average mass number. Most notably, the results of the Fly's Eye
group \cite{Gaisser-1993} indicate a rather heavy composition well above 
the knee. If taken at face value, their results represent a composition of
mainly very heavy nuclei, like iron, at $10^{17}$\,eV.

Experiments measuring the composition right at the knee of the
spectrum obtained either ambiguous or even conflicting results.
Results with no significant change
\cite{Zhu-1990,Ahlen-1992b} or a slight increase of heavy elements
\cite{Khristiansen-1994} have been reported. Other 
groups found more significant increases
of heavy elements \cite{Ren-1988a,Freudenreich-1990,Mitsui-1995} 
or, on the other hand, predominantly protons \cite{Cebula-1990}.
Further measurements are needed, if the cosmic-ray composition
should help to resolve the question where cosmic rays are accelerated.

The method proposed in this paper should be relatively easy to implement
at sites where an air-shower array with an angular resolution below
one degree already exists. The method is mainly a measurement of the
longitudinal shower development by the angles of muons with respect to
the shower axis. It does not require to measure many muons in a single 
shower because it uses only the {\em inclusive}
angular distributions, nor is accurate timing required.
Indeed, if several muons are measured in one event, they are treated
separately.

Not only the average muon angles but also the average electron angles 
with respect to the shower axis are sensitive to the primary
composition. Because the detector response 
is, in general, better understood for muon tracks than for electron tracks,
this paper focuses mainly on the muons.
Using the tracking detectors and the air-shower array, the
method can be supplemented by traditional methods like the
average $\mu/e$ ratio or the muon lateral distribution.
Although one can think of complex experiments dedicated
to measuring the cosmic-ray composition, like KASKADE \cite{Rebel-1993},
where more pieces of information are collected, the
purpose of this paper is to demonstrate that even 
the muon angles alone provide significant information about
the composition.

The proposed method relies on shower simulations to obtain
an absolute value of an average mass number, just like
other indirect methods. If suitable tracking detectors are used,
it has the advantage that essentially
all required detector parameters for the simulation can be
obtained from the measured data as a function of shower size.
Therefore, it can be particularly sensitive to changes in the composition
across the measured shower size spectrum, even if a comparison
with simulations using different interaction models may yield
different absolute values.

An analysis of data taken with ten CRT detectors 
-- tracking detectors of 2.5\,m$^2$ sensitive area each \cite{CRT-NIM-1} --
and the HEGRA air-shower array \cite{Fonseca-1995} on La Palma is
in progress \cite{future-CRT-results}. In the data of CRT and HEGRA
both muon and electron tracks are analysed. Although the intention of
this paper is to promote the method to other air-shower experiments,
the simulations shown in this paper are specific to the combination of the
CRT and HEGRA experiments. These simulations take into account the
response of both components -- the tracking detectors and the
air-shower array -- in much detail. After an outline of the method,
the specific simulations are described to demonstrate that the
method can be very well applied with existing detector technology.
Experimental requirements for application of the method and
limitations by shower simulations are shown in a more general context,
not specific to CRT and HEGRA.


\section{Outline of the method}

First measurements of the angles of muons with respect to the
corresponding air showers or, in alternative terms, of the apparent
height of origin of the muons were done in the 1960s \cite{Earnshaw-1973}.
At that time neither the angular resolutions of air-shower arrays
nor the accuracy of shower simulations were appropriate to use
such measurements for investigating the mass composition of
primary cosmic rays. For an early comparison of measured
and simulated height of muon origin see \cite{Gaisser-1978}.

The proposed method is based on the fact that showers initiated
by heavy primaries, like iron nuclei, develop on average
at smaller atmospheric depths, i.e.\ higher altitudes, than proton showers.
Muons from showers of heavy primaries arrive, therefore, at smaller
angles with respect to the shower axis for any given core distance.
The proposed method is not intended to measure individual primaries
but to measure accurately an average number of the composition, like 
$\langle\ln A\rangle$, the average logarithm of the primary mass numbers.
Measuring many tracks in a single shower is, therefore, not necessary
and only helps to improve statistics.

\begin{figure}[htbp]
\epsfxsize=0.6\textwidth
\hc{\epsffile{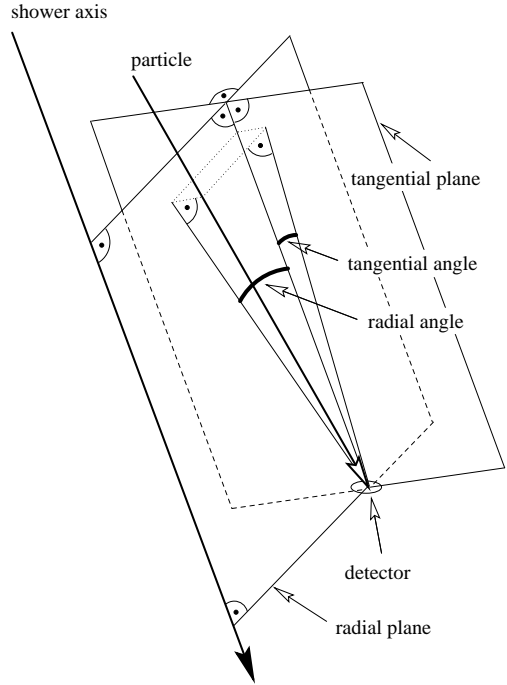}}
\caption[Geometry of radial and tangential angles]{The
geometry of radial and tangential angles between the
shower axis, e.g.\ as reconstructed from air-shower data,
and the measured track of a particle.}
\label{fig:geometry}
\end{figure}

The angles of muons or other particles with respect to the shower axis are 
best expressed in terms of their {\em radial angle\/} and {\em tangential angle\/}
components. The radial angle is the projection onto the plane defined
by the geometry of the shower axis and the location of the muon detector.
Throughout this paper, a negative radial angle is assigned to
a particle flying away from the shower axis. In fact, at large
core distances most particles have such a negative radial angle.
The tangential angle (also called the transverse angle) is the projection
onto the plane perpendicular to the `radial' plane and parallel to the
shower axis (see figure~\ref{fig:geometry}). 
For almost vertical
showers a more convenient definition with projections into vertical
planes defined only by the positions of the shower core and the
detector is essentially equivalent. 

\begin{figure}[htbp]
\epsfxsize=0.6\textwidth
\hc{\epsffile{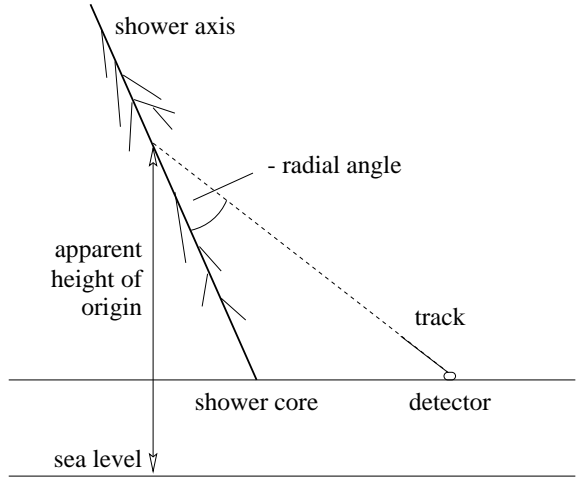}}
\caption[Apparent height of origin]{An apparent
height of origin can be derived from the track projected onto
the radial plane.}
\label{fig:geo-height}
\end{figure}

The radial projection can be
used to derive an {\em apparent height of origin\/} of a particle
(figure~\ref{fig:geo-height}). Although radial angle and apparent
height of origin are almost equivalent, the radial angle is the
more robust variable.
This is due to the fact that
a small scattering or experimental smearing around zero radial
angle corresponds to $\pm\infty$ apparent height.
In addition, the derived height for each muon depends directly on the
resolution of the core position. In particular, the last point turns out 
to be a major drawback of an analysis in terms of the apparent height.
The core position resolution is usually not very precisely known
as a function of shower size, zenith angle, and position within the
array. An analysis in terms of the apparent height is, for this
reason, limited to core distances much larger than the core position
resolution.

\begin{figure}[htbp]
\epsfxsize=0.8\textwidth
\hc{\epsffile{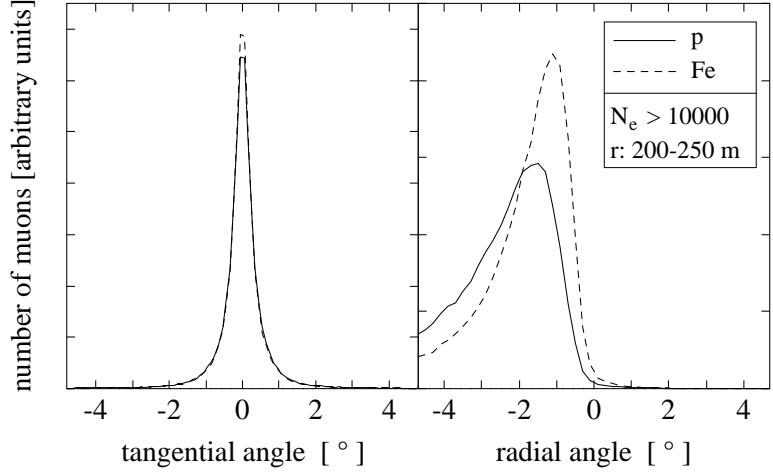}}
\caption[Distributions of tangential and radial muon angles]{Distributions
of tangential and radial angles of shower muons in simulated 
proton and iron showers, respectively,
with $N_e>10000$, at core distances of 200--250~m. 
Perfect angular resolution but a realistic detection
efficiency as appropriate for CRT detectors is assumed.}
\label{fig:tang+rad}
\end{figure}

The distribution of tangential angles (see figure \ref{fig:tang+rad})
is symmetric and its width is
dominated by multiple scattering and the lateral distribution of the
parent particle generation. For the muons in UHE air showers the
intrinsic distribution is very narrow, apart from tails due
to some low-energy muons. This distribution shows
little difference between proton- and iron-initiated showers,
even for an ideal angular resolution.
Under certain experimental constraints the tangential angle
can be used to measure the combined angular resolution of
air-shower array and tracking detectors.

\begin{figure}[htbp]
\epsfxsize=0.8\textwidth
\hc{\epsffile{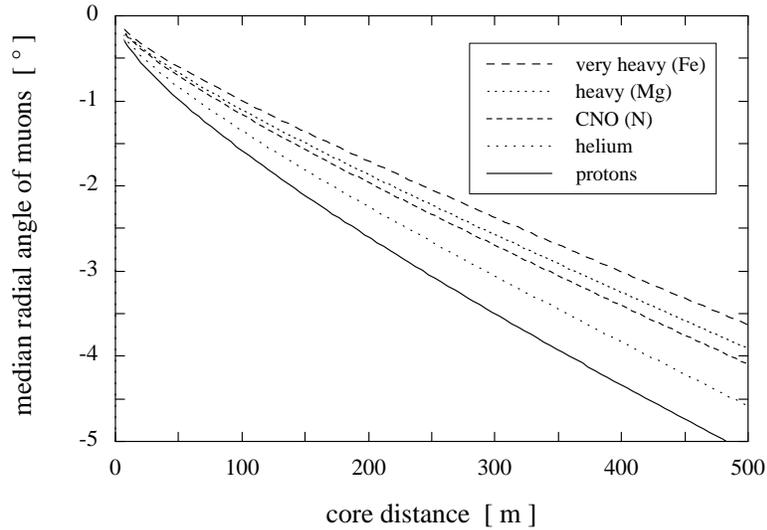}}
\caption[Radial angle of muons for different primaries]{Median radial angle of 
muons in showers initiated by different primaries, for a minimum shower size of 
10$^4$ and angular resolutions and detector responses as 
appropriate for CRT detectors and the HEGRA scintillator array 
(see section \ref{sect:shower+det-sim}).}
\label{fig:rad-ang-primaries}
\end{figure}

The radial angle, representing the longitudinal shower development,
is the quantity sensitive to the composition. The {\em average}
(which may be mean, median, or mode) radial angle depends in a
characteristic way on the core distance. 
To compare the simulated average radial-angle curves
with measurements requires to know the combined angular resolution of the
air-shower array and tracking detectors and to some extent also the
core position resolution of the array. However, the average radial angle
at any given core distance changes very little with shower size or zenith angle.
The figures in this paper are intended as examples. They are based on
simulations using the CORSIKA code (see section \ref{sect:shower+det-sim}).
The simulations assume an experimental resolution as observed
for CRT detectors at the HEGRA array \cite{CRT-NIM-2}. 
See figure~\ref{fig:rad-ang-primaries}
for the median radial angles of different primary types.
At any core distance the average radial angle is almost a linear function
of $\langle\ln A\rangle$, where $A$ is the mass number of the primary.

Although a similar difference between showers initiated by protons or
iron nuclei is also present in the radial-angle curves of electrons,
the electron component is less usefull for composition studies than muons.
First of all, a very detailed simulation of the behaviour of
the tracking detectors to electrons, gammas, muons, and hadrons 
is required to compare with experimental data. 
For muons the detector is generally much better understood,
although punch-throughs of other particles have to be considered.

The second reason in favour of muons is based on the shower selection
introduced by the air-shower array. To achieve the same shower size
($N_e$) at ground level, showers initiated by heavy primaries require
a larger primary energy than proton showers. Due to the steep spectrum,
conventional air-shower arrays will, therefore, mainly see proton showers
at any shower size, even if protons account for only half of the
mass composition at a given energy per particle. Muons can almost
compensate for this bias because, at the same shower size, a proton
shower contains less muons than a shower of a heavy primary.

Because of the compensation of the selection bias by the larger number
of muons for heavy primaries the quantity
\begin{equation}
\Lambda_\mu = \frac{1}{\sum w_i}\sum_i \,w_i\,\frac{\langle\alpha_i\rangle -
  \langle\alpha_{i,{\rm p}}\rangle}{\langle\alpha_{i,{\rm Fe}}\rangle -
  \langle\alpha_{i,{\rm p}}\rangle}
\end{equation}
is even for a mixed composition essentially proportional to $\langle\ln A\rangle$:
\begin{equation}
   \langle\ln A \rangle \approx \Lambda_\mu\,\ln 56.
\label{eq:lnA-Lambda}
\end{equation}
In this context $\langle\alpha_i\rangle$ is the measured (or simulated) 
median radial angle in the core distance interval $i$, 
$\langle\alpha_{i,{\rm p}}\rangle$ and $\langle\alpha_{i,{\rm Fe}}\rangle$
are the expected median values for pure protons and pure iron 
nuclei from the simulations, and the weights $w_i$ take
the statistical accuracy of measured as well as simulated average values
into account. In the simplest possible formulation, ignoring statistical
errors of the simulation and correlations of errors in different
radial bins and taking into account only $\sigma(\langle\alpha_i\rangle)$
as the measured accuracy of $\langle\alpha_i\rangle$,
\begin{equation}
   w_i = \Bigl( \sigma(\langle\alpha_i\rangle) \;/\; 
  ({\langle\alpha_{i,{\rm Fe}}\rangle -
  \langle\alpha_{i,{\rm p}}\rangle}) \Bigr)^{-2}.
\end{equation}

For compositions as reported by direct measurements, equation
\ref{eq:lnA-Lambda} 
holds to
better than 0.05 at all investigated shower size intervals --
at least for CRT and HEGRA to which the simulations
correspond. Although this relation is quite coincidental,
it is expected to hold as well for most other existing
air-shower arrays in combination with tracking detectors
of a low energy threshold for muons.
In the case of extreme compositions like
pure helium, however, the relation may be wrong by up to 0.26
in terms of $\langle\ln A\rangle$.

\begin{figure}[htbp]
\epsfxsize=0.8\textwidth
\hc{\epsffile{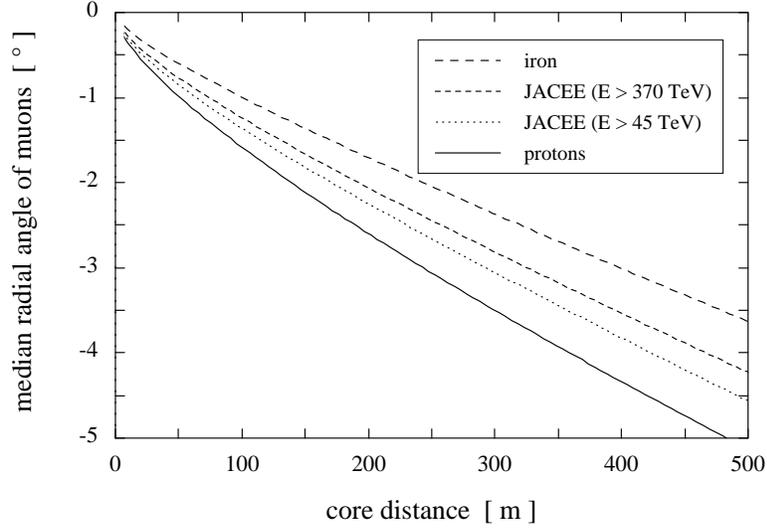}}
\caption[Radial angle of muons for different compositions]{Median radial 
angle of muons in showers
for different energy-independent model compositions (pure protons, pure iron,
and two compositions as measured by the JACEE collaboration 
\cite{Asakimori-1993ab}).
Simulations for $N_e>10^4$, as in figure \ref{fig:rad-ang-primaries}.}
\label{fig:mu-rad-ang-composite}
\end{figure}

\begin{figure}[htbp]
\epsfxsize=0.8\textwidth
\hc{\epsffile{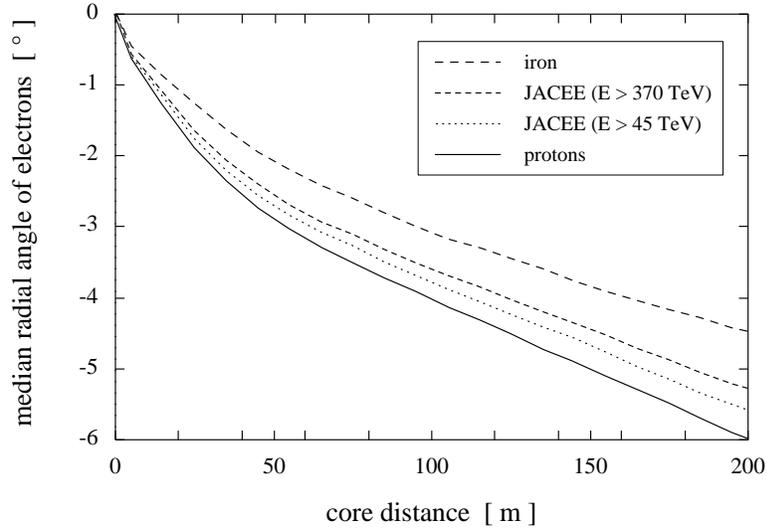}}
\caption[Radial angle of electrons for different compositions]{Simulated 
median radial angle of particles identified as electrons,
for the same model compositions as in figure~\ref{fig:mu-rad-ang-composite}.
$N_e>10^4$, as in figure \ref{fig:rad-ang-primaries}.}
\label{fig:elec-rad-ang-composite}
\end{figure}

Figure \ref{fig:mu-rad-ang-composite} shows that the median (or mean) radial
angle of muons is indeed sensitive to changes in the composition
as indicated by direct measurements \cite{Asakimori-1993ab} below
the knee. Even the electron radial angles can be used for that
purpose (see figure \ref{fig:elec-rad-ang-composite}) but
a more accurate knowledge of the detector response is required in that case.
In addition, $\Lambda_e$ (defined in analogy to $\Lambda_\mu$), is not
proportional to $\langle\ln A\rangle$ but is more sensitive to the
fraction of light nuclei than to heavy nuclei.


\section{Shower and detector simulations}
\label{sect:shower+det-sim}

The scope of this paper is beyond the measurements
carried out with CRT detectors and the HEGRA array.
Although the simulations presented here serve only as examples,
experimental effects of CRT detectors and the
HEG\-RA scintillator array are carefully taken into account. It may
be instructive to see which of these effects turn out to be
the most relevant ones.

The CORSIKA simulation code \cite{CORSIKA-1992,Knapp-1993}
was used in version 4.068 for all shower simulations presented in this paper.
Except for tests of the accuracy of the shower simulations (see section
\ref{sect:shower-accuracy}) the VENUS hadron-nucleus and nucleus-nucleus
interaction code \cite{Werner-1993} was used for high-energy
hadronic interactions, GHEISHA \cite{Fesefeldt-1985} for 
hadronic interactions below 80~GeV, and EGS4 \cite{Nelson-1985} for
the electromagnetic component -- all in the framework of CORSIKA.
The simulations take also the geomagnetic field at La Palma into account.
A total of 5580 showers (1860 proton showers and 930 showers
for each of the other four types of primaries, He, N, Mg, and Fe) were generated 
isotropically in the zenith angle range 0$^\circ$ to 32$^\circ$.
The energy ranges were selected to cover approximately the same
shower sizes, from 20~TeV to 2~PeV for protons increasing to
40~TeV to 4~PeV for iron nuclei. Showers were generated with a
$E^{-1.7}$ differential energy spectrum to have sufficient
numbers of showers at all energies and weighted by $E^{-1}$ for the
analysis. The analysis was restricted to shower sizes where
contributions from showers outside the simulated energy ranges
are negligible. 

Concerning the CRT detectors \cite{CRT-NIM-1}, it should be 
mentioned that muons are identified as pairs of tracks in two
drift chambers, one above and one below a 10~cm thick iron plate.
The measured angles between the two tracks of an identified muon are
required to be less than 2.5$^\circ$ in two perpendicular projections. 
Electrons are identified as non-muon tracks in the upper drift chamber.
Results of very detailed simulations of the response of CRT detectors to various
types of particles (electrons, gammas, muons, pions, and protons) -- as
outlined in \cite{CRT-NIM-2} and described in detail in \cite{Zink-1995} --
and the measured detector performance \cite{CRT-NIM-2}
were carefully parametrized. This includes detection efficiencies
and angular resolutions as functions of particle energies, zenith angles,
and track densities -- independent
for each particle type and for the identification as an electron track 
and as a muon track. As a consequence, the simulations include
not only punch-through of electrons but also of hadrons.
It should be noted that the energy dependence for
detecting genuine muons is fully described by multiple
scattering and energy loss in the iron plate \cite{CRT-NIM-2}.
CRT detectors are mainly sensitive to muons above about
1~GeV. Electrons, on the other hand, are detected above some 10~MeV.
The effect of multiple scattering in the detector container
and the iron plate are included in the angular resolutions.

For a detailed simulation of the HEGRA array see \cite{Martinez-1995}.
For the simulations presented here, a much simpler approach is
sufficient because the simulations are restricted to shower
sizes where the HEGRA array is fully efficient, independent of
shower age, core position, and zenith angle.
The resolutions for shower sizes and core positions as obtained
by the detailed simulations were parametrized. 
The HEGRA angular resolution is implemented in the simulations
in such a way that the combined
angular resolution of CRT and HEGRA -- as measured by the muon
tangential angles as a function of shower size and zenith angle 
\cite{CRT-NIM-2} -- are fully reproduced.

It may be instructive to note that the median radial-angle curves
presented in this paper are not significantly altered by
modifying the assumed angular resolution
within the experimental limits. Modifying the assumed
core position resolution has an impact on the median radial
angles only within a few ten meters from the core. 
Punch-throughs, although included in some detail, turned out
to be of minor importance with CRT detectors, due to their
excellent punch-through rejection. The most important
experimental parameters in the simulations are the energy 
dependence of the detection efficiencies and the combined angular resolution
of tracking detectors and air-shower array. In the case of CRT detectors
and the HEGRA array both are very well known for muons and even
sufficiently well known for electrons. 

It appears very reasonable that the response of other types of 
tracking detectors and other air-shower arrays can also be 
understood well enough to take advantage of the radial-angle method. 
However, a few experimental requirements as outlined in section
\ref{sect:requirements} should be observed for that purpose.


\section{Accuracy of shower simulations}
\label{sect:shower-accuracy}

Apart from the detector response, a major uncertainty in the
interpretation of any indirect measurement of the
cosmic-ray mass composition is due to the limited accuracy of
the shower simulations. Despite much progress in recent
years, the nucleus-nucleus and hadron-nucleus interaction models in 
air-shower simulations lead to non-negligible theoretical
uncertainties. The CORSIKA simulation code
attempts a particularly detailed and accurate shower simulation. 
In the version 4.068, as used for this paper, it offers several
independent interaction models.

\begin{figure}[htbp]
\epsfxsize=0.8\textwidth
\hc{\epsffile{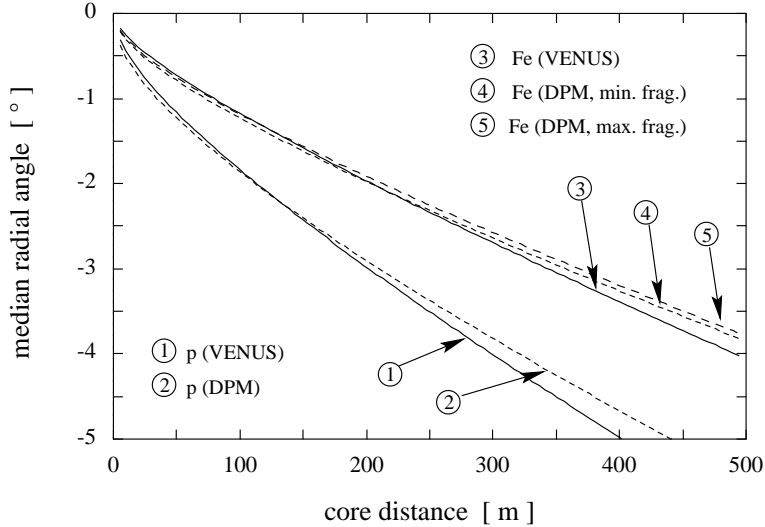}}
\caption[Radial angle of muons for interaction models]{Median radial 
angle of muons in proton and iron showers with different interaction
models (10000\,$<$\,$N_e$\,$<$\,20000, with fixed height of first interaction
to keep shower-to-shower fluctuations small). 1: protons with
VENUS/GHEISHA, 2: protons with DPM, 3: iron with VENUS/GHEISHA,
4(5): iron with DPM and no (full) fragmentation of the spectators
in projectile nuclei.}
\label{fig:rad-ang-syserr}
\end{figure}

Results for different
interaction models, using the VENUS and GHEI\-SHA 
options on one side and DPM on the other side, and for
different fragmentation (evaporation) of the projectile nuclei have
been compared in the energy range $10^{14}$--$10^{15}$\,eV.
Systematic differences in the average shower sizes of up
to 20\% are found and also differences in the
shower fluctuations. If the median radial-angle curves are compared
in the same ranges of shower sizes, the differences between models are
very small within core distances up to about 200\,m, increasing to
about 10\% of the proton-iron difference at a core distance of 400\,m 
(figure~\ref{fig:rad-ang-syserr}).
That increase of systematic uncertainties is due to the fact that 
hard interactions resulting in high-$p_t$ secondaries are not yet
included very accurately.
With all interaction models the shower-size dependence of the
radial-angle curve is small. A 30\% systematic error in the
relation of simulated to measured shower sizes would result
only in a systematic error of the median radial angles of 3\% of
the proton-iron difference.
Moreover, this dependence is essentially 
the same for different models.

Another uncertainty in the longitudinal shower development is
due to the sparse particle physics data in the very forward rapidity 
region at high energies. Until more such data is available and 
incorporated into the interaction models it seems desirable to have also
indirect composition measurements in an energy range overlapping with 
direct measurements. When searching only for a change of the
composition near the knee of the cosmic-ray spectrum, the poor
knowledge of the very forward region should not be a limitation.
Different parametrizations of parton densities which pose problems
for shower simulations at extremely high energies \cite{Capdevielle-1995}
are not of concern near $10^{15}$\,eV primary energy.


\section{Experimental requirements}
\label{sect:requirements}

To exploit the described method for composition studies several
experimental requirements should be met. These are requirements for
the {\em particle identification} with a tracking detector and 
the {\em resolution} of angles and positions.

Because the particles of the electromagnetic shower component, 
$e^\pm$ and $\gamma$, have angular distributions much different from
those of muons, a good electron rejection is important. Hadrons
only matter very close to the shower core. Due to the broad
muon lateral distribution it is generally not such a problem
to identify muons at large core distances but more of a problem
to reject random, non-shower muons. This can be easily solved
by a timing information of a few hundred nanoseconds accuracy
and by the angular correlation of shower muons with the shower axis.
Another requirement for the
muon identification is due to the fact that the radial angles
are well correlated with the muon momentum, with higher-energy
muons usually coming from further up in the atmosphere and being
better aligned with the shower axis. The muon identification should,
therefore, be well understood also as a function of momentum and
zenith angle.
For a high muon-energy threshold, a very good angular resolution
is required, while for a low threshold the multiple scattering
and magnetic deflection of muons cannot be neglected.

The combined angular resolution of air-shower array
and tracking detector should be better than about 1$^\circ$
in each projection but has to be known as a function of shower size
to compare measurements with simulations. Definitely sufficient would be
a combined resolution of some 0.5$^\circ$ if the muon-energy 
threshold is as low as about 1~GeV.
This would be difficult to achieve with a tracking detector
entirely below thick shielding material which also scatters the muons.
A good uniformity of the resolution over azimuth angles would
be of advantage for the important detector calibration. Only with a
uniform resolution can the angular resolution be derived from
the tangential angle distribution of the same data set which is
used to compare the radial angles with simulations. Otherwise,
shower-size dependent systematic errors can arise. 

The alignment of the tracking detectors with respect to the
reference frame as defined by showers reconstructed from array data
should be known to better than about 0.1$^\circ$. This may require some
careful calibrations but is certainly feasible. 

The core position resolution which also enters into the simulations 
has to be determined from the air-shower array alone. Keeping 
selection effects of the air-shower array as small as possible
demands to select only showers well contained in the array
and with shower sizes well above the trigger and reconstruction
thresholds. For many of the existing air-shower arrays that
would correspond to threshold energies of some 80--200\,TeV,
just below where direct measurements are running out of statistics.
Reasonable errors in the shower size and energy calibrations of the array 
do not compromise the method as long as the same experimental resolution
is used in the simulations as seen in the data to be compared with
these simulations.

With the 25~m$^2$ sensitive area of the CRT detectors the
statistics with data of about one year would be the limit for 
primary energies above several $10^{16}$~eV. A few hundred
square meters area would be sufficient
in order to achieve
some overlap with the Fly's Eye composition measurements.

All these requirements can be fulfilled with present
technology for large-area tracking detectors and existing air-shower
arrays. Although simultaneous precise (nanosecond) timing
information of the muons could supplement the described method
(see for example \cite{Danilova-1994}),
precise timing is not required and might be difficult to accomplish
with large-area detectors.


\section{Concluding remarks}

The average radial angle of particles are sensitive to the
cosmic-ray mass composition. Muons have, among other particles in air
showers, many advantages. Their intrinsic angular distribution is
very narrow and, thus, the tangential-angle distribution can be
used to calibrate the radial-angle resolution as a function of
shower size. Muons can be distinguished from other particle types
very well. The larger number of muons in showers initiated by 
heavy primaries compensates for the $N_e$ shower selection bias
which favours light primaries.

Like any other
indirect measurement of some average of the cosmic-ray composition,
that method requires comparison of measured data with simulations.
Compared to many other indirect methods, the muon radial-angle method
has two major advantages: First, the most important detector effects
to be included in the simulation can be measured very well and, second,
systematic errors, for example due to the interaction model,
can be easily checked by comparing measurements at low energies
(a few hundred TeV) with simulations for directly measured
compositions as in \cite{Asakimori-1993ab}. 

Despite some systematic uncertainties in the shower simulations
an average mass number ($\langle\ln A\rangle$) can be derived
and compared with direct measurements well below the knee. 
Regardless of that, the muon radial-angle method can be very
sensitive to changes in the composition. The radial-angle
method would not require a very large experimental effort if
existing air-shower arrays are supplemented with suitable
tracking detectors.




\begin{thebibliography}{10}
\itemsep=0pt

\bibitem{Asakimori-1993ab}
K.~Asakimori et~al.,
\newblock in: Proc. 23rd Intern. Cosmic Ray Conf., Vol.\ 2, pp.\ 21 and 25, 1993.

\bibitem{Ichimura-1993b}
M.~Ichimura et~al.,
\newblock Phys. Rev. D\relax 48 (1993) 1949.

\bibitem{Gaisser-1993}
T.~K. Gaisser et~al.,
\newblock Phys. Rev. D\relax 47 (1993) 1919.

\bibitem{Zhu-1990}
{Zhu Qingqi} et~al.,
\newblock J. Phys. G\,\relax 16 (1990) 295.

\bibitem{Ahlen-1992b}
S.~Ahlen et~al.,
\newblock Phys. Rev. D\relax 46 (1992) 895.

\bibitem{Khristiansen-1994}
G.~B. Khristiansen et~al.,
\newblock Astroparticle Physics 2 (1994) 127.

\bibitem{Ren-1988a}
J.~R. Ren et~al.,
\newblock Phys. Rev. D\relax 38 (1988) 1404.

\bibitem{Freudenreich-1990}
H.~T. Freudenreich et~al.,
\newblock Phys. Rev. D\relax 41 (1990) 2732.

\bibitem{Mitsui-1995}
K.~Mitsui et~al.,
\newblock Astroparticle Physics 3 (1995) 125.

\bibitem{Cebula-1990}
D.~Cebula et~al.,
\newblock Astroph. J. 358 (1990) 637.

\bibitem{Rebel-1993}
H.~Rebel et~al.,
\newblock in: Jones \cite{ann-arbor-1992}, p. 575.

\bibitem{CRT-NIM-1}
K.~Bernl{\"o}hr et~al.,
\newblock Nucl. Instr. and Meth. A369 (Jan. 1996), in press
\newblock (Paper I).

\bibitem{Fonseca-1995}
V.~Fonseca,
\newblock in: Currents in High-Energy Astrophysics, eds.\ M.~M. Shapiro,
  R.~Silberberg and J.~P. Wefel 
  (NATO ASI Series Vol. 458, Kluwer, Dordrecht, 1995) p. 143.

\bibitem{future-CRT-results}
K.~Bernl{\"o}hr et~al.,
\newblock in preparation.

\bibitem{Earnshaw-1973}
J.~C. Earnshaw et~al.,
\newblock J. Phys. A 6 (1973) 1244.

\bibitem{Gaisser-1978}
T.~K. Gaisser et~al.,
\newblock Rev. Mod. Phys. 50 (1978) 859.

\bibitem{CRT-NIM-2}
K.~Bernl{\"o}hr et~al.,
\newblock Nucl. Instr. and Meth. A369 (Jan. 1996), in press
\newblock (Paper II).

\bibitem{CORSIKA-1992}
J.~N. Capdevielle et~al.,
\newblock Technical Report KfK 4998, 
  Forschungszentrum Karlsruhe, 1992.

\bibitem{Knapp-1993}
J.~N. Capdevielle et~al.,
\newblock in: Jones \cite{ann-arbor-1992}, p. 545.

\bibitem{Werner-1993}
K.~Werner,
\newblock Phys. Rep. 232 (1993) 87.

\bibitem{Fesefeldt-1985}
H.~Fesefeldt,
\newblock Report PI-THA 85/02, RWTH Aachen, 1985.

\bibitem{Nelson-1985}
W.~R. Nelson et~al.,
\newblock SLAC Report 265, Stanford Linear Accelerator Center,
  1985.

\bibitem{Zink-1995}
R.~Zink,
\newblock PhD thesis, 1995.

\bibitem{Martinez-1995}
S.~Martinez et~al.,
\newblock Nucl. Instr. and Meth. A357 (1995) 567.

\bibitem{Capdevielle-1995}
J.~N. Capdevielle and R.~Attalah,
\newblock J. Phys. G\,\relax 21 (1995) 121.

\bibitem{Danilova-1994}
T.~V. Danilova et~al.,
\newblock J. Phys. G\,\relax 20 (1994) 961.

\bibitem{ann-arbor-1992}
L.~Jones, editor,
\newblock Very High Energy Cosmic-Ray Interactions, 
  {VIIth} International Symposium, Ann Arbor, MI, USA, 1993, 
  (AIP Conf. Proc. 276, University of Michigan, 1993).

\end{thebibliography}
\end{document}